\documentclass{article}
\usepackage{jinstpub} % for details on the use of the package, please see the JINST-author-manual
\usepackage{lineno}
\usepackage[utf8]{inputenc}
\usepackage{graphicx}
\usepackage[pdf]{graphviz}
\usepackage{siunitx}
\usepackage{eqparbox}
\DeclareSIUnit \vrms {\ensuremath{\mathrm{V_{rms}}}}
\DeclareSIUnit \vpp {\ensuremath{\mathrm{V_{pp}}}}
\DeclareSIUnit \barn {b}

\usepackage{hyperref}
\usepackage{lineno}
\title{Comparison of readout systems for high-rate silicon photo-multiplier applications}

\author[a]{M.~L.~Wong,\thanks{Corresponding author.} }% orcid.org/0000-0002-9530-1222
\author[a, b]{M.~Ko\l{}odziej,\thanks{Corresponding author.} }% orcid.org/0000-0001-8016-949
\author[c]{K.~Briggl,}% orcid.org/0000-0003-0274-8023
\author[d]{R.~Hetzel,}% orcid.org/0000-0002-6142-6609
\author[a]{G.~Korcyl,}% orcid.org/0000-0001-6244-7287
\author[a]{R.~Lalik,}% orcid.org/0000-0003-1313-3729
\author[e]{A.~Malige,}%orcid.org/0000-0002-5855-2372
\author[a]{A.~Magiera,}% orcid.org/0000-0002-7561-6366
\author[a]{G.~Ostrzo\l{}ek,}% orcid.org/0009-0007-3108-7910
\author[a]{K.~Rusiecka,} % orcid.org/0000-0003-2554-0612
\author[d]{A.~Stahl,} % orcid.org/0000-0002-8369-7506
\author[a]{V.~Urbanevych,}% orcid.org/0000-0002-1858-4708
\author[f]{M.~Wiebusch,}% orcid.org/0000-0003-0381-9022
\author[a]{and A.~Wro\'nska} % orcid.org/0000-0003-0126-3315

\affiliation[a]{M. Smoluchowski Institute of Physics, Jagiellonian University, Kraków, Poland}
\affiliation[b]{Doctoral School of Exact and Natural Sciences, Jagiellonian University, Krak\'ow, Poland}
\affiliation[c]{Kirchhoff-Institute of Physics, Universit\"{a}t Heidelberg, Heidelberg, Germany}
\affiliation[d]{Physics Institute III B, RWTH Aachen University, Aachen, Germany}
\affiliation[e]{Department of Physics, Columbia University, USA}
\affiliation[f]{GSI, Darmstadt, Germany}
\emailAdd{ming-liang.wong@clermont.in2p3.fr}
\emailAdd{mkolodziej@doctoral.uj.edu.pl}

\abstract{
Recent years have shown an increased use of silicon photo-multipliers (SiPM) in experiments as they are of reasonable cost, have relatively low power consumption and are easily available in a variety of form factors allowing for a large number of readout channels. At the same time, experiments are generating data at increasingly high rates requiring the use of more efficient readout systems. In this work, the dead time, efficiency, dynamic range, coincidence time resolution and energy resolution of five different readout systems at various stages of maturity are evaluated to determine the best system for acquiring data from a detector in a high rate experiment. Additional functionalities of the systems are also discussed.
}


\begin{document}
\maketitle
\flushbottom
\section{Introduction}

\begin{figure}
    \centering
    \includegraphics[width=0.7\textwidth]{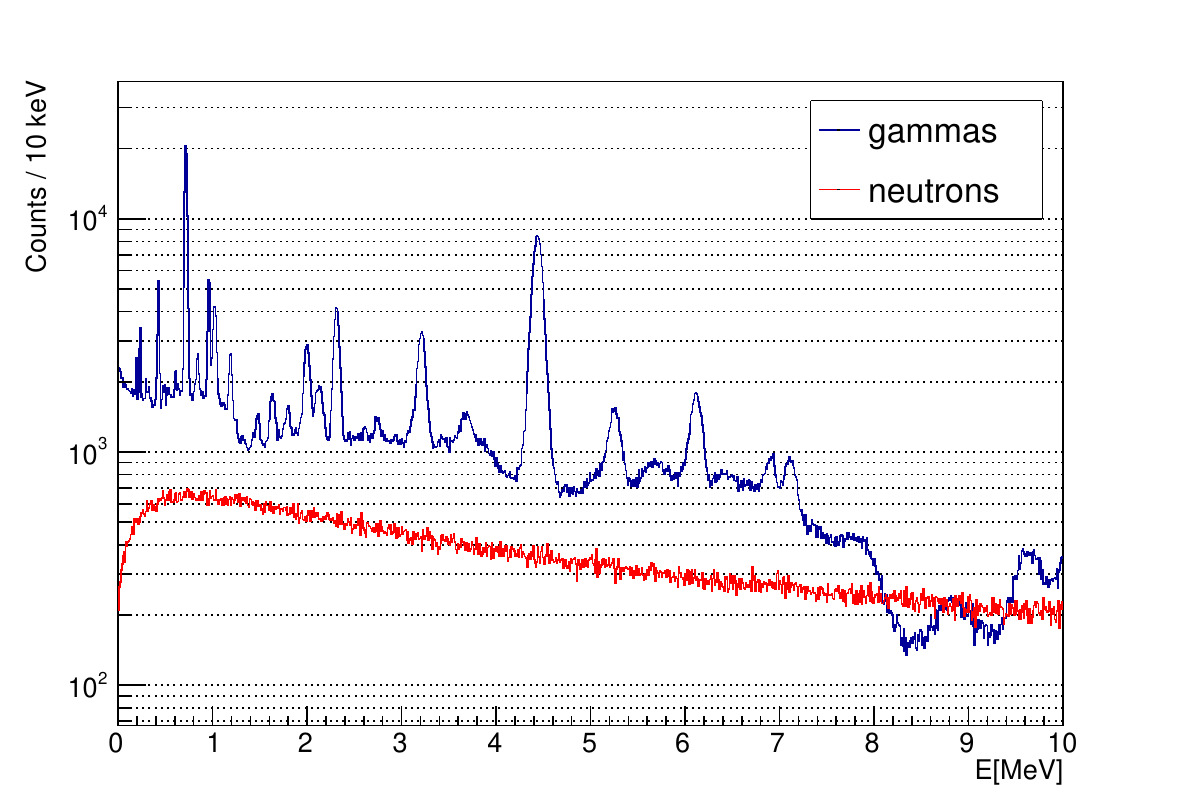}
    \caption{Spectra of gammas and neutrons created in a PMMA target within the SiFi-CC geometrical acceptance. The total numbers of gammas and neutrons are about the same order of magnitude.}
    \label{fig:photons_neutrons_sifi}
\end{figure}

High-rate readout and digitization of data from accelerator experiments are now becoming commonplace. The High-Luminosity LHC project aims to increase its instantaneous luminosity by a factor of five beyond the original design value to \SI{5e-34}{\per\centi\meter\squared\per\second} and the annual integrated luminosity to \SI{250}{\per\femto\barn} which is a factor of ten from the original design value~\cite{ZurbanoFernandez:2020cco}. Consequently, the detectors in LHC are also upgrading the readout electronics to match the increase in luminosity~\cite{Aad:2022noz, LaRosa:2021cof, LHCb:2023hlw}. The intensity frontier experiments, e.g. Belle II \cite{Yamada2015}, Mu2e \cite{Mu2e:2014fns}, Muon g-2 \cite{Gohn2015} and others require electronics and readout systems that can filter and process raw data at high efficiency and low dead time. For example, the Belle II detector at design luminosity is expected to deliver a raw data rate of \SI{2}{GB/s} \cite{Kuhr:2019jra} and they use advanced triggering schemes to reduce this to below \SI{30}{kHz}. The Mu2e experiment is expecting an average incoming data rate of \SI{35}{GB/s} and uses triggering to reject uninteresting data. The Muon g-2 experiment has an incoming data rate of \SI{20}{Gbit/s} \cite{Khaw:2017cvx} and uses GPUs to pre-select pulses before processing by the DAQ. 
Accelerators are used not only in basic research experiments, but also in applications such as proton radiotherapy. There, large rate capability is also a key factor as the statistics is strictly limited by the maximum allowable radiation dose. Secondary radiation emitted during proton radiotherapy such as prompt gammas can be used for proton range monitoring. Some of the techniques are prompt gamma spectroscopy (PGS) requiring excellent energy resolution, prompt gamma timing (PGT) where timing resolution is crucial, and for prompt gamma imaging (PGI) both are essential. One of the detector types considered for the latter method is a Compton camera. Several groups, among others \cite{BABIANO2020163228, Munoz2021, Draeger2018,Koide2018}, are developing such detectors. A Monte Carlo study with the Silicon photomultipiers and scintillating Fibers based Compton Camera (SiFi-CC) \cite{KASPER2020317} estimates event rates reaching the SiFi-CC detector to be about \SI{43}{Mcps} considering PG and neutron events as shown in Figure~\ref{fig:photons_neutrons_sifi}. A readout system of a CC suitable for the application in proton therapy, as well as for other applications where high rates are involved, should fulfill the following criteria:
\renewcommand{\theenumi}{\roman{enumi}}%
\begin{enumerate}
    \item Short dead time (of the order of \si{\micro\second} or smaller).
    \item Activity in one channel not causing any dead time in the other channels, allowing for independent processing.
    \item Programmable trigger logic (channel-to-channel coincidences).
    \item Scalability (up to $\mathcal{O}(10^3$) channels).
\end{enumerate}
In this study, several readout systems fulfilling these criteria at least partially were compared by means of their dead time, efficiency, energy- and coincidence timing resolutions. Their additional features were also discussed.

\section{Readout systems}
The overview of SiPM readout systems reported in 2012~\cite{delaTaille:2012} covered mostly ASICs in an early development phase, and not available off-the-shelf. However, recent progress in the field resulted in many different readout systems that can be used for SiPM-based detectors, e.g. calorimeters, scintillation fiber trackers, and Compton cameras. The requirements for such a system suitable for operation in a high-rate environment are low dead time and adequate dynamic range capable of measuring energy deposits. For PGI, the dynamic range should extend up to \SI{7}{MeV}. We examined five readout systems, among which four are scalable and one was used as a reference. The five systems are chosen based on ease of procurement and measurement readiness. This includes documentation and access to their designers. These systems also pass early screening by meeting high rate operational requirements which can be quickly discerned from their data sheets or designers.
Some of the tested systems provide pulse digitization per channel, which proves useful in decreasing the overall dead time. Individual systems are presented in detail in the following sections, and their main properties are summarized in Table~\ref{tab:readout_summary}.
\begin{table}
    \centering
    \begin{tabular}{c|c|c|c|c|c|c}
    \hline
        System           & Type & Channel \#  & ADC [bit] & Max power/     & Input signal & Price per\\
                         &   & per unit &  & channel [mW] & polarity & channel [EUR]\\
        \hline
        DT5742           & ASIC & 8        & 12       & 140         & +/- & -\\
        TwinPeaks+TRB5sc & Board & 16        & 8        & <100        & +/-& 57\\
        A5202            & ASIC & 64       & 13       & 141         & +& 70\\
        KLauS6b          & ASIC & 36       & 10, 12   & 3.6         & + & 20\\
        TOFPET2c         & ASIC & 64       & 10       & 5           & + & 17\\
        \hline
    \end{tabular}
    \caption{Properties of the various readout systems. TwinPeaks is not available as an ASIC. The A5202 is a board consisting of two Citiroc-1A chips. Price per channel is estimated for a system with 6000 channels as of 2021.}
    \label{tab:readout_summary}
\end{table}
\subsection{DT5742 - CAEN Desktop Digitizer}
The CAEN Desktop Digitizer is part of a family of stand-alone instruments housing a high-speed multichannel flash 12-bit ADC with local memory and FPGA for real-time data processing. The digitizer is based on the Switched Capacitor DT5742~\cite{4774700} chip. The DT5742 is a waveform sampler using a series of 1024 capacitors (analog memory) in which the analog input signal is continuously sampled circularly. The sampling frequency is \SI{5}{GHz} by default and with options for \SI{2.5}{GHz}, \SI{1}{GHz}, and \SI{750}{MHz}. In this study, the digitizer operates with \SI{1}{GHz}. In the desktop form factor, it is a convenient tool with acquisition software that can be operated on Linux. It enables full waveform analysis of the registered pulses, enabling insights into the noise and pile-up issues. An earlier study of the DT5742 performance has been conducted in~\cite{Wang:2022bhf}. However, the system has limited possibilities of scaling up, so it is included in the comparison as a reference only.

\subsection{TwinPeaks+TRB5sc}
TwinPeaks is a custom analog add-on board designed for the DESPEC experiment at GSI/FAIR to facilitate fast-timing measurements of photo-multiplier tube pulses from the FAst TIMing Array detector setup (FATIMA~\cite{RUDIGIER2020163967}). In this study, the TwinPeaks add-on board~\cite{BANERJEE2022166357} is connected to the front-end readout board TRB5sc~\cite{TRBfamily} instead of originally to the TAMEX4~\cite{10.1117/12.2567973} board. We only have access to the TRB5sc, but both are designed by the Experimentelektronik (EEL) group at GSI/FAIR Darmstadt and could interface with TwinPeaks without further modifications. Multiple TwinPeaks+TRB5sc can be linked to a single TRB3~\cite{Neiser:2013yma} motherboard, which incorporates a trigger system capable of sending a timing signal to all connected boards, collects data registered by them and transmits the data to the event building software. A single TwinPeaks board accommodates up to 16 analog inputs while the TRB5sc provides precise time measurement and TrbNet connectivity – an interconnection protocol between the members of the TRB hardware family. Various scale systems can be constructed in a tree-like architecture with one master module and multiple hubs to which many TRB5scs with TwinPeaks addons can be connected. TRB3 Together with TRB5sc and TwinPeaks create an FPGA-based system for SiPM readout. The energy spectrum as registered by the detector is directly converted into a pulse width spectrum, thus eliminating the need for a sampling ADC and digital waveform processing. Consequently, the energy information can be recorded by a TDC up to \SI{1.2}{\micro\second} after the arrival of the pulse. The system is mature but only a limited number of boards are available for use in the DESPEC experiment. Its custom acquisition software, DABC~\cite{Adamczewski-Musch:2015arx} as well as the network protocol, TrbNet~\cite{Michel:2010ffa} are both required for the system to work on Linux.

\subsection{A5202}
The A5202~\cite{A5202_Manual_2023} is a part of a multi-board system by CAEN, designed for the readout of multichannel detectors. The unit houses two Citiroc-1A~\cite{DatasheetCitiroc1A} chips by Weeroc, 32 channels each. It can operate in spectroscopy, counting, timing and time stamped spectroscopy modes. The spectroscopy and time stamped spectroscopy modes are the only ones with precise pulse height information, necessary for our purpose. However, as the 13-bit ADC runs serially for all channels, the dead time due to conversion with Citiroc-1A is relatively high, of about \SI{10}{\micro\second}. In this study, efficiency and dynamic range measurements were performed in the spectroscopy mode, while for the remaining ones, where time information was also required, time stamped spectroscopy mode was utilized. These two modes perform very similarly, except that for the latter one, the time stamp information is also stored. The A5202 is a mature, commercially available system with accompanying software for data acquisition on Windows OS. It features tunable gain using high- (HG) or low-gain (LG) preamplifiers, the latter was used in the measurements described here. The system can be scaled up by using a concentrator board DT5215~\cite{CAEN_concentrator_board}, which can link up to 128 A5202 boards with a total number of 8192 channels. If DT5215 is unavailable, up to 16 A5202 can be daisy-chained. That board is then connected to a DAQ computer through Ethernet or USB~3.0. Furthermore, multiple DT5215 boards can be synchronized in order to further extend the total number of channels. All the components are parts of the FERS-5200 system~\cite{fers5200} by CAEN Technologies Inc.

\subsection{KLauS6b}
The board was originally designed by a group at the Heidelberg University for CALICE Analogue Hadronic Calorimeter (AHCAL)~\cite{Laudrain:2022fmh} \& ScECAL~\cite{CALICE:2013zlb} detectors targeting low gain SiPMs of charge range between 15 to \SI{150}{pC}. It can also operate with four different gain modes, with the lowest one spanning the dynamic range up to \SI{450}{pC}. KLauS6b~\cite{Briggl_2014} is an auto/external-triggered, gain-configurable 36-channel ASIC fabricated in UMC \SI{180}{nm} CMOS technology. Analog data from the 36 channels are digitized and sent off-chip using either an I2C interface (up to \SI{20}{Mbit/s}) or a faster \SI{160}{Mbit/s} LVDS interface, providing a 8b10b encoded data stream. The device that communicates with KLauS6b via I2C fetches data until the board is empty, and bundles all received hits into one acquisition object. KLauS6b samples the peak voltage and then converts it to digital code word by an ADC. A 10-bit or 12-bit ADC are available for one ASIC according to the purposes; in this work, we chose to work with a 10-bit ADC. The ADC block allows for conversion times in the order of \SI{400}{ns} with a \SI{40}{MHz} internal system clock, not including the sampling time of typically \SI{90}{ns}. The system is still under development requiring an additional Raspberry Pi~4 connected to the development board and only a limited number of ASICs are available. It comes with acquisition software that works on Linux.

\subsection{TOFPET2c}
This is a readout and digitization 64-channel ASIC featuring low-noise and low-power for time-of-flight (TOF) measurements and other applications using SiPMs~\cite{DIFRANCESCO2016194}. The digitization modules are present for each readout channel. The chip has quad-buffered TDCs and QDCs in each channel. The chip is part of a complete, scalable readout platform offered as an off-the-shelf solution by PETsys Electronics~\cite{FEBD}. Up to 16 ASICs can be connected to a single front-end board~\cite{FEBD} that has a clock frequency of \SI{200}{MHz} resulting in a total of 1024 channels. Groups of ASICs can be easily programmed to select coincident events within a pre-set time window. A combination of using a PCIe with 3~SFP+ optical/copper connectors connecting to chains of a FEB/D and Trig\&Clk boards can transmit data up to \SI{6.6}{Gbit/s}. This is a mature, commercially available product ready to be used off the shelf with acquisition software on Linux.

\section{Test setups}
 
\begin{figure}
\centering
\includegraphics[width = 0.7\textwidth]{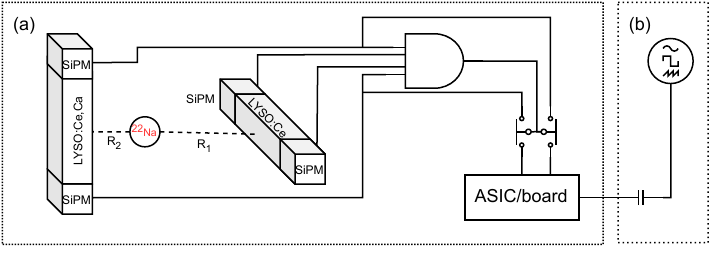}
\caption{Detection test setups used in the study. The details vary slightly between different readout systems. $R_1\approx61.4$~mm is the distance from the source to the reference detector and $R_2\approx30.5$~mm is the distance from the source to the middle of the LYSO:Ce,Ca fiber. Setup (a) is used for dead time, coincidence time and energy resolution measurements, and setup (b) for efficiency and dynamic range studies.}
\label{fig:setup}
\end{figure}

The DAQ systems were characterized with both a pulse generator and a detector setup consisting of scintillators with SiPM readout.The sensitive part of the detector setup (Figure~\ref{fig:setup}a) has a form of a fiber stack made of 64 $1.28 \times1.28 \times $\SI{100}{mm}$^3$ LYSO:Ce,Ca fibers, manufactured by Taiwan Applied Crystals. The stack has four layers and each fiber is wrapped in Al foil. For the other setup (Figure~\ref{fig:setup}b), a 81160A Pulse Function Arbitrary Noise Generator~\cite{PulseGenerator} along with a 33 pF or 600 pF capacitor, depending on the readout system.
However, only one fiber was exploited in the tests. There is a variety of possible setup architectures in real world applications, differing in number of channels by orders of magnitude, so it is most practical to consider the smallest unit available for evaluation. This least biased method minimizes the differences due to the different hardware design decisions, e.g. ADC, buffers. With this in mind, the measured channel properties could be scaled up effectively by the reader when evaluating larger systems.

Such choice of the scintillation crystal and the wrapping material provided the best trade-off between the energy- (7.73\% at \SI{511}{keV}) and position (\SI{33.4}{mm}) resolutions and was decided to be the optimal solution for the SiFi-CC setup~\cite{rusiecka2023sificc}. On each end of the fiber stack, there was an array of 64 $1\times1\mathrm{mm}^2$ PM1125-WB Ketek\footnote{Ketek SiPMs are now manufactured by Broadcom} SiPMs~\cite{broadcom:AFBR-S4K11C0125B}. These SiPMs have 1600 microcells and a fast recovery time of \SI{30}{ns}. They were operated at an overvoltage of \SI{4.5}{V}. The SiPMs and fibers are coupled with custom-made, \SI{0.5}{mm} thick Elastosil RT~604 optical pads and aligned centrally to each other, the total pitch between subsequent elements in the stack or array being \SI{1.36}{mm}. For coincidence time- and energy-resolution measurements, a $^{22}$Na source is placed between the electronic collimator and the detector system to irradiate only a \SI{3}{mm} segment in the middle of the test fiber in the detector, the electronic collimation procedure was as described in \cite{Anfre2007}. The $^{22}$Na source had an activity of about \SI{1.7}{MBq} at the time of the measurements. The reference detector consists of a $3\times3\times100$ mm$^3$ LYSO:Ce fiber with 3x3 Hamamatsu S13360-3050CS SiPMs on both sides. The coincidence trigger is constructed with the CAEN V792 module from the binary AND operation of 4 discriminated SiPM pulses from both sides of the reference detector, as well as from two ends of the fiber stack. This NIM trigger signal is then used as an input for external triggering in the various readout systems. This trigger scheme reduces the LYSO:Ce,Ca background.

\section{Methodology}
\label{sec:Methodology}

\subsection{Dead time}
\begin{figure}
    \centering
    \includegraphics[width=\textwidth]{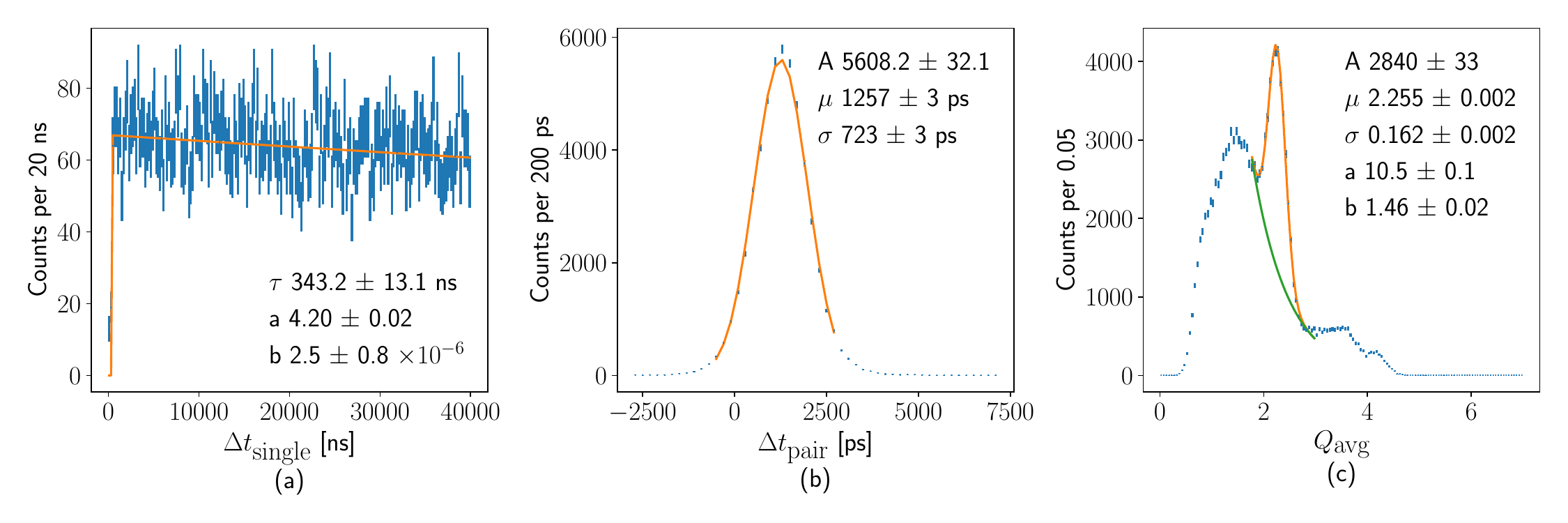}
    \caption{Determination of various performance metrics shown on the example of the TOFPET2c system data. (a) Time interval between subsequent events in a channel. The average time interval is \SI{419}{\micro\second}. (b) Arrival time difference between two channels on opposite ends of a fiber. (c) $^{22}$Na spectrum from a single LYSO:Ce,Ca fiber, constructed as a geometric mean of the signals registered on its two ends.}
    \label{fig:methodology}
\end{figure}

The technique used to determine the system dead time $\tau$ requires a single channel and a radioactive source. Besides the signals generated by the interactions of the $^{22}$Na gammas with the detector material, there was also a contribution from gammas of the LYSO:Ce,Ca intrinsic activity, which added to the occupation of the readout channels. The readout systems can register the hit times and from this obtain the time interval between subsequent hits $\Delta t$. The distribution of time intervals between the hits is Poissonian and thus can be modelled as an exponential, whose exponent is rate-dependent. 
However, due to dead time, a gap near zero of the width $\tau$  appears, due to events missed because of digitization time, pile up and the ASIC being in BUSY mode. To quantify the dead time $\tau$, a piece-wise function is fit to the $\Delta t$ distribution:
\begin{equation}
    y=\begin{cases}
        0 & \text{if } t<\tau, \\
        e^{a-b(t-\tau) } & \text{otherwise}.
    \end{cases}        
\end{equation}
An example of such an analysis for the TOFPET2c data is shown in Figure~\ref{fig:methodology}a, where the determined dead time was  $\tau= \SI{343.2}{ns}$.

\subsection{Coincidence time resolution}
The requirement for having a narrow coincidence time between two correlated signals is an effective way to reduce both electronic noise and the physical background. In our case, the two SiPMs located on opposite ends of a single LYSO:Ce,Ca fiber should receive hits within some time window, typically less than \SI{3}{ns} given the length of the crystal of \SI{100}{mm}. In order to mitigate the walk effect on the timing resolution, the time difference between the two SiPMs is constructed on an event sample containing only fully absorbed annihilation photons of \SI{511}{keV}, i.e. with energy depositions within $\pm3\sigma$ from the mean annihilation peak energy. Figure~\ref{fig:methodology}b shows a sample histogram of the time difference, with the cut on energy deposit. The coincidence time resolution is then extracted as the standard deviation of the Gaussian fit to the data. The errors are calculated as $\sqrt{N}$, where N is the number of counts.

\subsection{Energy resolution}
\label{sec:enres}
The back-to-back production of the annihilation gamma from the $\beta^+$ decay of $^{22}$Na is a good physics filter that reduces background. Hence, the \SI{511}{keV} peak is useful to evaluate the energy resolution $R$, defined as
\begin{equation}
\label{eq_enres}
    R = \frac{\sigma}{\mu},
\end{equation}
where $\sigma$ and $\mu$ are the standard deviation and the peak position of a Gaussian fitted to describe the photo peak. Energy resolution defined in Equation~\ref{eq_enres} incorporates several effects: SiPM parameters, scintillating fiber properties like light yield or intrinsic resolution. However, those were the same for all examined systems, as only the DAQ system was changed between measurements. Thus, such a method remains a valid comparison of DAQ systems. For our data, an exponential function and a Gaussian are chosen to locally describe the background and signal shape. An example of such an energy deposit spectrum is shown in Figure~\ref{fig:methodology}c as the geometric mean of the signals registered on both fiber ends, $Q_\mathrm{avg} = \sqrt{Q_LQ_R}$. This quantity is independent of the hit position along the fiber \cite{DaZhi2004}, which is not the case for individual signals $Q_L$ and $Q_R$.

\subsection{Efficiency}
In order to determine the readout efficiency for a single channel, SiPM signals are emulated by injecting square pulses from a pulse generator passing through a capacitor into one of the test channels at increasing frequencies at a fixed amplitude $A = \SI{300}{mV}$. The A5202 typically digitizes data from all 64 channels serially, so for this study all but one channel are active (other channels are masked). However, it should be noted that this will produce higher efficiencies compared to operations with more channels. Spectroscopy mode measurements with the A5202 are done with low gain, LG=5. The readout efficiency, $\epsilon$ is the ratio of the measured and the true (injected) rate $r$.
The efficiency curve follows the shape of a sigmoid
\begin{equation}
    \epsilon = \frac{1}{1 + e^{a+b(r-r_0) } },
\end{equation}
where $r_0$ is the mid-point of the falloff and $a$, $b$ the exponential shape parameters. The true rate at which the efficiency drops to 90\% is used to evaluate the different systems.
% {\color{red} $m=n/(1+n\tau)$ and $m=n\theta/(e^{\theta n \tau}+\theta-1)$ for fitting the real vs measured rates do not work. They provide additional insight into the source of dead times. Is the dead time extendable/non-extendable? ADCs by themselves are non-extendable however, we have in combination the SiPMs.}

\subsection{Dynamic range}
Readout systems are optimized for the signal amplitudes and charges typical for certain applications and given by energy deposits and the characteristics of the detectors under test. An RC high-pass filter is used to emulate SiPM pulses with a falling time constant of \SI{30}{ns} using square pulses with a peak voltage, $\Delta V$ and a capacitor with capacitance, $C$=\SI{600}{pF} capacitor for TOFPET2c and TwinPeaks+TRB5sc and \SI{33}{pF} for KLauS6b and DT5742 realizing the total injected charge of $Q_\mathrm{in} = C\Delta V$. For A5202, an additional voltage dividing circuit was necessary to reduce the pedestal. The internal feedback capacitance of the low gain preamplifier, $C_f$ is set to \SI{150}{fF} providing an amplification factor of 10. The signal magnitude is expressed in ADC, QDC or TOT units, depending on the readout system. The signal is plotted against the injected charge, $Q_\mathrm{in}$. The charge conversion factor is obtained as the slope in the linear region of this plot. This linear region determines the dynamic range of the system. The linear region is determined by first checking the linear fit slope uncertainties to not exceed 3\%, otherwise the fit range is reduced. Then, a fifth order polynomial is used to find the limits where the deviation from the linear fit is more than 10\%. Typically, the lower limit is at the pedestal. If no deviation from linearity was observed, the full range was considered to be the system's dynamic range.
\section{Results}
\begin{figure}[ht]
    \centering
    \includegraphics[width=0.5\textwidth]{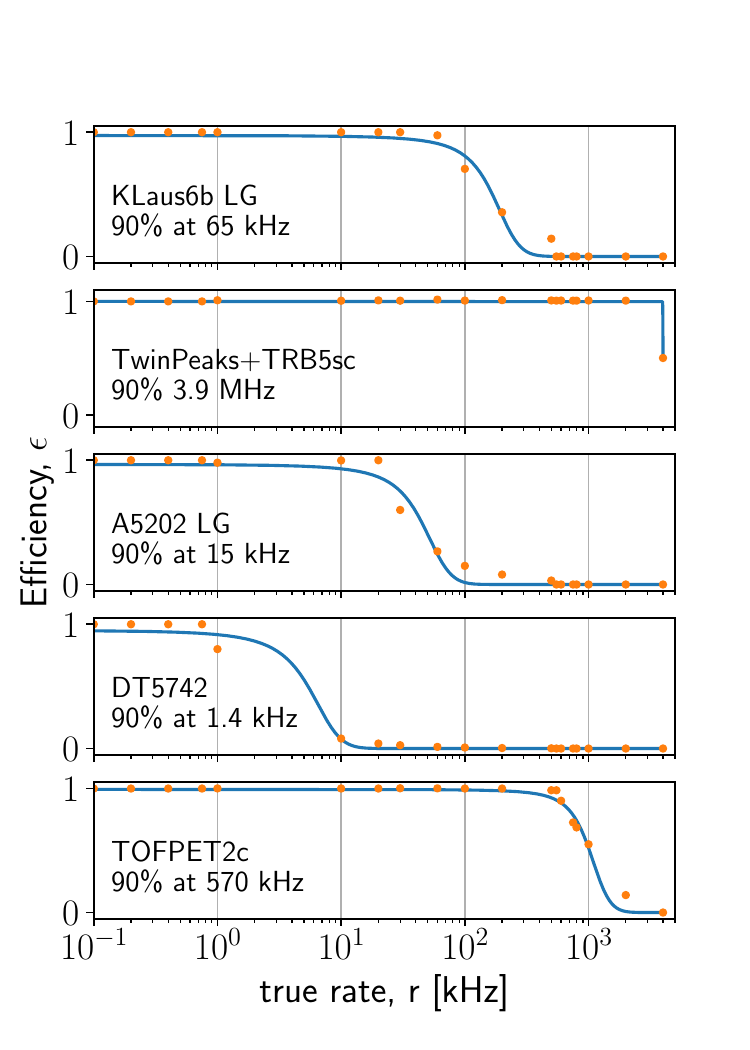}
    \caption{Comparison of the readout systems efficiencies up to \SI{4}{MHz}.}
    \label{fig:efficiency}
\end{figure}

\begin{figure}
    \centering
    \includegraphics[width=\textwidth]{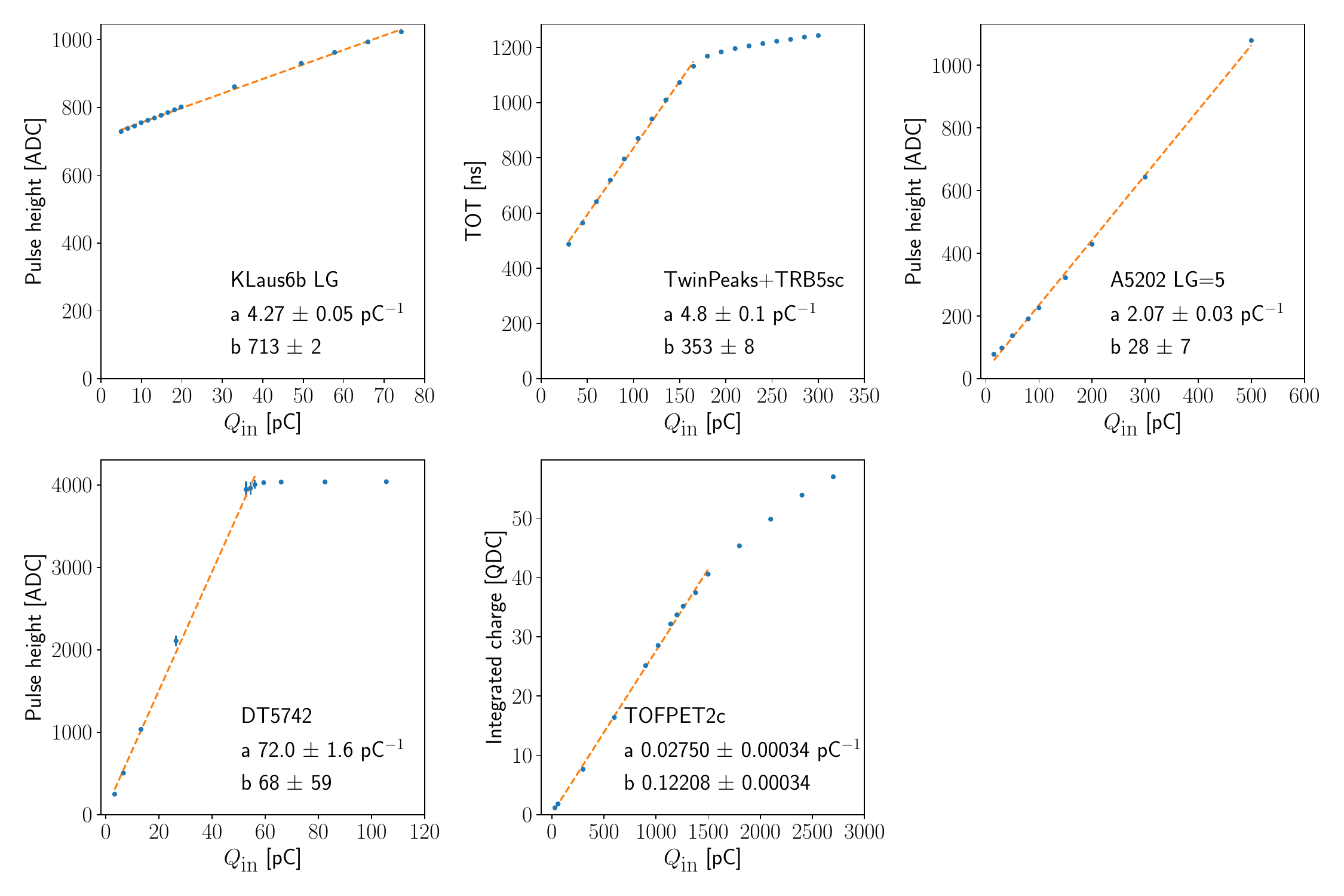}
    \caption{Comparison of the readout systems' dynamic range. The linear regions are indicated and fit with a dotted line where the charge conversion factor is $a$ and the pedestal, $b$.}
    \label{fig:dynamicrange}
\end{figure}

The properties described in Section~\ref{sec:Methodology} for all the investigated readout systems are collected in Tables~\ref{tab:measurements_comparison},~\ref{tab:measurements_comparison_pulse_generator} and illustrated in Figures~\ref{fig:efficiency},~\ref{fig:dynamicrange}.
\begin{description}
    \item[Dead time] TOFPET2c has the lowest dead time of \SI{0.343(13)}{\micro\second} due to the quad-buffered TDCs and charge integration QDCs in each channel. KLauS6b has comparably low dead time (see Table~\ref{tab:measurements_comparison}) of \SI{0.352(77)}{\micro\s} thanks to digitization available per channel. TwinPeaks+TRB5sc does not require ADC, so it is also relatively fast, the dead time being about twice as large as the winner in this category. The A5202 simultaneously acquires and digitizes from all 64 channels when triggered, thus having about 100 times larger dead time than the winner. For the DT5742, which also features a simultaneous digitization and saves full waveforms, the measured dead time is 1000 times larger compared to KLauS6b and TOFPET2c.
    \item[Coincidence time resolution] The system with the best coincidence time resolution (see Table~\ref{tab:measurements_comparison}) is TOFPET2c with \SI{0.723(3)}{ns}. DT5742 performs slightly worse with \SI{1.152(1)}{ns}, the next one is KLauS6b with \SI{1.53(7)}{ns}, followed by A5202 and TwinPeaks+TRB5sc (3 and 10 times worse than the winner, respectively).
    \item[Energy resolution] The best energy resolution (see Table~\ref{tab:measurements_comparison}) was obtained with TOFPET2c which is 7.2(1)\%, followed by the DT5742 (20\% worse), A5202 (30\% worse), TwinPeaks+TRB5sc (60\% worse) and finally by KLauS6b (75\% worse).
    \item[Efficiency] The efficiency vs rate dependence in Figure~\ref{fig:efficiency} and Table~\ref{tab:measurements_comparison_pulse_generator} show that the TwinPeaks+TRB5sc system could operate with full efficiency at the highest rate amongst the investigated systems, which is \SI{3.9}{MHz}. This stems from the fact that TwinPeaks+TRB5sc does not perform time-costly analog-to-digital conversion. TOFPET2c ranks second in maintaining full efficiency at 7 times lower rate, followed by KLauS6b (60 times lower rate than the winner). Both A5202 and DT5742 have relatively low efficiency of the order of 1-\SI{10}{kHz}.
    \item[Dynamic range] TOFPET2c covers a significantly larger dynamic range than any other investigated system (Figure~\ref{fig:dynamicrange}, Table~\ref{tab:measurements_comparison_pulse_generator}, which is up to \SI{1899}{pC}. The second largest dynamic range (4 times less than the TOFPET2c's) is attributed to A5202. It is followed by TwinPeaks+TRB5sc (9 times less than TOFPET2c) and later by KLauS6b and DT5742 (25 times less than TOFPET2c).  In case of TOFPET2c, non-linearity correction function was applied to the data, as suggested in the documentation~\cite{TOFPETSoftwareUserGuide}. KLauS6b and A5202 did not reach the nonlinear region as it is limited by their ADCs. The systems have very different dynamic ranges as they have been optimized for different SiPM applications.
\end{description}

\begin{table}
    \centering
    \begin{tabular}{r|c|c|c}
    \hline
        System       &  Dead time [\SI{}{\micro\second}]    & Coincidence time resolution [ns]  & Energy resolution [\%]\\
        \hline
      
       KLauS6b LG     & \eqmakebox[cint][r]{0.352(77)}          & \eqmakebox[cint][r]{1.53(7)}           & \eqmakebox[cint][r]{12.4(1)} \\
        TwinPeaks+TRB5sc    & \eqmakebox[cint][r]{0.870(9)}           & \eqmakebox[cint][r]{10.5(3)}              & \eqmakebox[cint][r]{11.1(13)} \\
        A5202 LG        & \eqmakebox[cint][r]{44(4)}        & \eqmakebox[cint][r]{3.0(2)}             & \eqmakebox[cint][r]{9.4(3)} \\
         DT5742       & \eqmakebox[cint][r]{443(97)}       & \eqmakebox[cint][r]{1.152(1)}           & \eqmakebox[cint][r]{8.55(4)} \\
        TOFPET2c     & \eqmakebox[cint][r]{0.343(13)}       & \eqmakebox[cint][r]{0.723(3)}             & \eqmakebox[cint][r]{7.2(1)}\\
        \hline
    \end{tabular}
    \caption{Comparison of the dead time, coincidence time resolution and energy resolution (as defined in Section~\ref{sec:enres}).
    % \text{*}For A5202, periodic structures observed in time difference spectrum were observed which caused problems in the fitting procedure. Thus for A5202, instead of fitting, we define the dead time value as the centre of the first populated bin and assume the uncertainty to be ...
    }
    \label{tab:measurements_comparison}
\end{table}
\begin{table}
    \centering
    
    \begin{tabular}{r|c|c|c}
    \hline
        System       &  Rate at 90\% efficiency [kHz]   & Dynamic range [pC] & Charge conversion factor [pC$^{-1}$]\\
        \hline
        KLauS6b LG     & \eqmakebox[cint][r]{65}          & \eqmakebox[cint][r]{74.25$^{\dag}$}           & \eqmakebox[cint][r]{4.27(5)}\\
        TwinPeaks+TRB5sc    & \eqmakebox[cint][r]{3900}           & \eqmakebox[cint][r]{202}              & \eqmakebox[cint][r]{4.8(1)}\\
        A5202 LG        & \eqmakebox[cint][r]{15}         & \eqmakebox[cint][r]{500$^{\dag}$}             & \eqmakebox[cint][r]{2.07(3)}\\
        DT5742       & \eqmakebox[cint][r]{1.4}       & \eqmakebox[cint][r]{62}           & \eqmakebox[cint][r]{72.0(1.6)}\\
        TOFPET2c     & \eqmakebox[cint][r]{570}       & \eqmakebox[cint][r]{1899}             & \eqmakebox[cint][r]{0.025(1)}\\
        \hline
    \end{tabular}
    \caption{Comparison of efficiency, dynamic range, charge conversion factor and pedestal for all examined detection systems. The capacitor tolerance of 20\% dominates the uncertainty. Dynamic ranges for which the deviation from linearity was not observed are marked with a \dag. The charge conversion factor unit is $\mathrm{x}/\mathrm{pC}$, where $\mathrm{x}$ denotes arbitrary ADC, QDC or TOT units depending on the system.} \label{tab:measurements_comparison_pulse_generator}
\end{table}

\section{Summary}
Five readout systems that are either commercially available or in active development have been tested. The combined results from the tests show that TOFPET2c is the best performing system for our application given its marginally lowest dead time of \SI{0.343(13)}{\micro\second}, widest linearity range of up to \SI{1899}{pC}, best coincidence time resolution of \SI{0.723(3)}{ns}, and best energy resolution at 7.2(1)\%. Additional benefits of TOFPET2c are ease of electronics configuration, lowest price per channel of all the compared systems (see Table~\ref{tab:readout_summary}) and a fully operational, bundled acquisition software. It should be noted that even though the A5202 did not obtain competitive results in our comparison apart from the second-best dynamic range, it is well suited for measurements requiring excellent photopeak separation at low intensities and not too high rates. Moreover, the channel-to-channel coincidence can be programmed in any way needed, providing a vast reduction of background data. However, this system's price per channel is the highest of all the compared ones. With KLauS6b, a dead time of \SI{0.352(77)}{\micro\second} was obtained, making it the second-best choice for detectors demanding low dead time. This is the second most affordable system. TwinPeaks+TRB5sc can operate with the highest data throughput of all the studied systems, which makes it a suitable candidate for applications where the efficiency is crucial. It is also the only system of the scalable ones that can operate on signals of both polarities. Its price per channel is relatively high. DT5742 provides second-best time- and energy resolution. However, it has higher dead time and worse efficiency when operating at higher rates, as well as a relatively smaller dynamic range. This system is not suitable for detectors with large number of channels, but its ability to register full waveforms makes it a good choice for applications still in the detector development and troubleshooting. While TOFPET2c is the optimal system for the application in the SiFi-CC detector, each of the tested DAQ systems turned out to perform very well in at least one category, as a result of various features being prioritized by the producers. Depending on the use case, a different system can be optimal and by performing this comparison, we would like to provide a broad look on the data acquisition systems for high rate SiPM applications as well as to facilitate the choice of one.
\section*{Acknowledgements}
This work was supported by the Polish National Science Centre (grant 2017/26/E/ST2/00618) and by the Jagiellonian University (MNS2021 U1U/P05/NO/03.29 and RSM U1U/W17/NO/28).

%\bibliographystyle{JHEP}
%\bibliography{references.bib}

\providecommand{\href}[2]{#2}\begingroup\raggedright\endgroup
\end{document}